\RequirePackage{lineno}
\documentclass[%
superscriptaddress,
showpacs,
amsmath,amssymb,
aps,
prl,
twocolumn,
]{revtex4-1}

\usepackage{verbatim}
\usepackage{graphicx}
\usepackage{dcolumn}
\usepackage{bm}
\usepackage{hyperref}
\newcommand{\invfb}{\rm{fb}^{-1}}

\begin{document}
\title{Search for standard-model $Z$ and Higgs bosons decaying into a bottom-antibottom quark pair in proton-antiproton collisions at 1.96 TeV}
\affiliation{Institute of Physics, Academia Sinica, Taipei, Taiwan 11529, Republic of China}
\affiliation{Argonne National Laboratory, Argonne, Illinois 60439, USA}
\affiliation{University of Athens, 157 71 Athens, Greece}
\affiliation{Institut de Fisica d'Altes Energies, ICREA, Universitat Autonoma de Barcelona, E-08193, Bellaterra (Barcelona), Spain}
\affiliation{Baylor University, Waco, Texas 76798, USA}
\affiliation{Istituto Nazionale di Fisica Nucleare Bologna, \ensuremath{^{kk}}University of Bologna, I-40127 Bologna, Italy}
\affiliation{University of California, Davis, Davis, California 95616, USA}
\affiliation{University of California, Los Angeles, Los Angeles, California 90024, USA}
\affiliation{Instituto de Fisica de Cantabria, CSIC-University of Cantabria, 39005 Santander, Spain}
\affiliation{Carnegie Mellon University, Pittsburgh, Pennsylvania 15213, USA}
\affiliation{Enrico Fermi Institute, University of Chicago, Chicago, Illinois 60637, USA}
\affiliation{Comenius University, 842 48 Bratislava, Slovakia; Institute of Experimental Physics, 040 01 Kosice, Slovakia}
\affiliation{Joint Institute for Nuclear Research, RU-141980 Dubna, Russia}
\affiliation{Duke University, Durham, North Carolina 27708, USA}
\affiliation{Fermi National Accelerator Laboratory, Batavia, Illinois 60510, USA}
\affiliation{University of Florida, Gainesville, Florida 32611, USA}
\affiliation{Laboratori Nazionali di Frascati, Istituto Nazionale di Fisica Nucleare, I-00044 Frascati, Italy}
\affiliation{University of Geneva, CH-1211 Geneva 4, Switzerland}
\affiliation{Glasgow University, Glasgow G12 8QQ, United Kingdom}
\affiliation{Harvard University, Cambridge, Massachusetts 02138, USA}
\affiliation{Division of High Energy Physics, Department of Physics, University of Helsinki, FIN-00014, Helsinki, Finland; Helsinki Institute of Physics, FIN-00014, Helsinki, Finland}
\affiliation{University of Illinois, Urbana, Illinois 61801, USA}
\affiliation{The Johns Hopkins University, Baltimore, Maryland 21218, USA}
\affiliation{Institut f\"{u}r Experimentelle Kernphysik, Karlsruhe Institute of Technology, D-76131 Karlsruhe, Germany}
\affiliation{Center for High Energy Physics: Kyungpook National University, Daegu 702-701, Korea; Seoul National University, Seoul 151-742, Korea; Sungkyunkwan University, Suwon 440-746, Korea; Korea Institute of Science and Technology Information, Daejeon 305-806, Korea; Chonnam National University, Gwangju 500-757, Korea; Chonbuk National University, Jeonju 561-756, Korea; Ewha Womans University, Seoul, 120-750, Korea}
\affiliation{Ernest Orlando Lawrence Berkeley National Laboratory, Berkeley, California 94720, USA}
\affiliation{University of Liverpool, Liverpool L69 7ZE, United Kingdom}
\affiliation{University College London, London WC1E 6BT, United Kingdom}
\affiliation{Centro de Investigaciones Energeticas Medioambientales y Tecnologicas, E-28040 Madrid, Spain}
\affiliation{Massachusetts Institute of Technology, Cambridge, Massachusetts 02139, USA}
\affiliation{University of Michigan, Ann Arbor, Michigan 48109, USA}
\affiliation{Michigan State University, East Lansing, Michigan 48824, USA}
\affiliation{Institution for Theoretical and Experimental Physics, ITEP, Moscow 117259, Russia}
\affiliation{University of New Mexico, Albuquerque, New Mexico 87131, USA}
\affiliation{The Ohio State University, Columbus, Ohio 43210, USA}
\affiliation{Okayama University, Okayama 700-8530, Japan}
\affiliation{Osaka City University, Osaka 558-8585, Japan}
\affiliation{University of Oxford, Oxford OX1 3RH, United Kingdom}
\affiliation{Istituto Nazionale di Fisica Nucleare, Sezione di Padova, \ensuremath{^{ll}}University of Padova, I-35131 Padova, Italy}
\affiliation{University of Pennsylvania, Philadelphia, Pennsylvania 19104, USA}
\affiliation{Istituto Nazionale di Fisica Nucleare Pisa, \ensuremath{^{mm}}University of Pisa, \ensuremath{^{nn}}University of Siena, \ensuremath{^{oo}}Scuola Normale Superiore, I-56127 Pisa, Italy, \ensuremath{^{pp}}INFN Pavia, I-27100 Pavia, Italy, \ensuremath{^{qq}}University of Pavia, I-27100 Pavia, Italy}
\affiliation{University of Pittsburgh, Pittsburgh, Pennsylvania 15260, USA}
\affiliation{Purdue University, West Lafayette, Indiana 47907, USA}
\affiliation{University of Rochester, Rochester, New York 14627, USA}
\affiliation{The Rockefeller University, New York, New York 10065, USA}
\affiliation{Istituto Nazionale di Fisica Nucleare, Sezione di Roma 1, \ensuremath{^{rr}}Sapienza Universit\`{a} di Roma, I-00185 Roma, Italy}
\affiliation{Mitchell Institute for Fundamental Physics and Astronomy, Texas A\&M University, College Station, Texas 77843, USA}
\affiliation{Istituto Nazionale di Fisica Nucleare Trieste, \ensuremath{^{ss}}Gruppo Collegato di Udine, \ensuremath{^{tt}}University of Udine, I-33100 Udine, Italy, \ensuremath{^{uu}}University of Trieste, I-34127 Trieste, Italy}
\affiliation{University of Tsukuba, Tsukuba, Ibaraki 305, Japan}
\affiliation{Tufts University, Medford, Massachusetts 02155, USA}
\affiliation{Waseda University, Tokyo 169, Japan}
\affiliation{Wayne State University, Detroit, Michigan 48201, USA}
\affiliation{University of Wisconsin-Madison, Madison, Wisconsin 53706, USA}
\affiliation{Yale University, New Haven, Connecticut 06520, USA}

\author{T.~Aaltonen}
\affiliation{Division of High Energy Physics, Department of Physics, University of Helsinki, FIN-00014, Helsinki, Finland; Helsinki Institute of Physics, FIN-00014, Helsinki, Finland}
\author{S.~Amerio\ensuremath{^{ll}}}
\affiliation{Istituto Nazionale di Fisica Nucleare, Sezione di Padova, \ensuremath{^{ll}}University of Padova, I-35131 Padova, Italy}
\author{D.~Amidei}
\affiliation{University of Michigan, Ann Arbor, Michigan 48109, USA}
\author{A.~Anastassov\ensuremath{^{w}}}
\affiliation{Fermi National Accelerator Laboratory, Batavia, Illinois 60510, USA}
\author{A.~Annovi}
\affiliation{Laboratori Nazionali di Frascati, Istituto Nazionale di Fisica Nucleare, I-00044 Frascati, Italy}
\author{J.~Antos}
\affiliation{Comenius University, 842 48 Bratislava, Slovakia; Institute of Experimental Physics, 040 01 Kosice, Slovakia}
\author{G.~Apollinari}
\affiliation{Fermi National Accelerator Laboratory, Batavia, Illinois 60510, USA}
\author{J.A.~Appel}
\affiliation{Fermi National Accelerator Laboratory, Batavia, Illinois 60510, USA}
\author{T.~Arisawa}
\affiliation{Waseda University, Tokyo 169, Japan}
\author{A.~Artikov}
\affiliation{Joint Institute for Nuclear Research, RU-141980 Dubna, Russia}
% ...

\collaboration{CDF Collaboration}
\altaffiliation[With visitors from]{
\ensuremath{^{a}}University of British Columbia, Vancouver, BC V6T 1Z1, Canada,
\ensuremath{^{b}}Istituto Nazionale di Fisica Nucleare, Sezione di Cagliari, 09042 Monserrato (Cagliari), Italy,
\ensuremath{^{c}}University of California Irvine, Irvine, CA 92697, USA,
\ensuremath{^{d}}Institute of Physics, Academy of Sciences of the Czech Republic, 182~21, Czech Republic,
\ensuremath{^{e}}CERN, CH-1211 Geneva, Switzerland,
\ensuremath{^{f}}Cornell University, Ithaca, NY 14853, USA,
\ensuremath{^{g}}University of Cyprus, Nicosia CY-1678, Cyprus,
\ensuremath{^{h}}Office of Science, U.S. Department of Energy, Washington, DC 20585, USA,
\ensuremath{^{i}}University College Dublin, Dublin 4, Ireland,
\ensuremath{^{j}}ETH, 8092 Z\"{u}rich, Switzerland,
\ensuremath{^{k}}University of Fukui, Fukui City, Fukui Prefecture, Japan 910-0017,
\ensuremath{^{l}}Universidad Iberoamericana, Lomas de Santa Fe, M\'{e}xico, C.P. 01219, Distrito Federal,
\ensuremath{^{m}}University of Iowa, Iowa City, IA 52242, USA,
\ensuremath{^{n}}Kinki University, Higashi-Osaka City, Japan 577-8502,
\ensuremath{^{o}}Kansas State University, Manhattan, KS 66506, USA,
\ensuremath{^{p}}Brookhaven National Laboratory, Upton, NY 11973, USA,
\ensuremath{^{q}}Istituto Nazionale di Fisica Nucleare, Sezione di Lecce, Via Arnesano, I-73100 Lecce, Italy,
\ensuremath{^{r}}Queen Mary, University of London, London, E1 4NS, United Kingdom,
\ensuremath{^{s}}University of Melbourne, Victoria 3010, Australia,
\ensuremath{^{t}}Muons, Inc., Batavia, IL 60510, USA,
\ensuremath{^{u}}Nagasaki Institute of Applied Science, Nagasaki 851-0193, Japan,
\ensuremath{^{v}}National Research Nuclear University, Moscow 115409, Russia,
\ensuremath{^{w}}Northwestern University, Evanston, IL 60208, USA,
\ensuremath{^{x}}University of Notre Dame, Notre Dame, IN 46556, USA,
\ensuremath{^{y}}Universidad de Oviedo, E-33007 Oviedo, Spain,
\ensuremath{^{z}}CNRS-IN2P3, Paris, F-75205 France,
\ensuremath{^{aa}}Universidad Tecnica Federico Santa Maria, 110v Valparaiso, Chile,
\ensuremath{^{bb}}Sejong University, Seoul 143-747, Korea,
\ensuremath{^{cc}}The University of Jordan, Amman 11942, Jordan,
\ensuremath{^{dd}}Universite catholique de Louvain, 1348 Louvain-La-Neuve, Belgium,
\ensuremath{^{ee}}University of Z\"{u}rich, 8006 Z\"{u}rich, Switzerland,
\ensuremath{^{ff}}Massachusetts General Hospital, Boston, MA 02114 USA,
\ensuremath{^{gg}}Harvard Medical School, Boston, MA 02114 USA,
\ensuremath{^{hh}}Hampton University, Hampton, VA 23668, USA,
\ensuremath{^{ii}}Los Alamos National Laboratory, Los Alamos, NM 87544, USA,
\ensuremath{^{jj}}Universit\`{a} degli Studi di Napoli Federico II, I-80138 Napoli, Italy
}
\noaffiliation


\begin{abstract}

The Collider Detector at Fermilab collected a unique sample of jets originating from bottom-quark fragmentation ($b$-jets) by selecting online proton-antiproton ($p\bar{p}$) collisions with a 
vertex displaced from the $p\bar{p}$ interaction point, consistent with the decay of a bottom-quark hadron. 
This data set, collected at a center-of-mass energy of 1.96\,TeV, and corresponding to an integrated luminosity of $5.4~\invfb$, is used to measure the $Z$-boson production cross section times branching ratio into $b\bar{b}$. The number of $Z\rightarrow b\bar{b}$ events is determined by fitting the dijet-mass distribution, 
while constraining the dominant $b$-jet background, originating from QCD multijet events, with data. The result, $\sigma(p\bar{p} \rightarrow Z) \times \mathcal{B}(Z \rightarrow b\bar{b})= 1.11\pm 0.08(\text{stat}) \pm 0.14(\text{syst})~\text{nb}$, is the most precise 
measurement of this process, and is consistent with the standard-model prediction. 
The data set is also used to search for Higgs-boson production. No significant signal is expected in our data and the first upper limit on the 
cross section for the inclusive $p\bar p \rightarrow H\rightarrow b\bar b$ process at $\sqrt{s}=$1.96\,TeV is set, corresponding to 33 times the expected standard-model cross section, 
or $\sigma = 40.6$\,pb, at the 95\% confidence level. 
  
\end{abstract}


\maketitle

\section{I. Introduction}

This paper reports the most precise measurement of the $\sigma(p\bar{p} \rightarrow Z) \times \mathcal{B}(Z \rightarrow b\bar{b})$ production cross section times branching ratio and the first inclusive search for the 
$p\bar{p} \rightarrow H \rightarrow b\bar{b}$ process in proton-antiproton ($p\bar{p}$) collisions at a center-of-mass energy of 1.96 TeV. These results arise 
from the analysis of the data set collected by the Collider Detector at Fermilab (CDF) with an online requirement of one jet with a displaced vertex from the primary $p\bar{p}$ interaction vertex (secondary vertex), consistent with a 
$b$-hadron decay ($b$-jet). 

The identification of the $Z\to b\bar{b}$ decay is challenging at hadron colliders because of the overwhelming irreducible background from multijet production (as is predicted by quantum chromodynamics, QCD), but 
it represents a benchmark capability as it validates the experimental techniques used in searches for events with $b$ quarks in the final states. In addition, it can be used for a direct test of the calibration of the 
jet-energy scale (JES) for $b$-jets, defined as a factor that measures the discrepancy between the effect of detector response and energy corrections in real and simulated $b$-jets. Finally, a large sample of 
$Z\to b\bar{b}$ events, combined with an algorithm for the identification of the charge of the $b$ quarks, allows the measurement of the $b\bar{b}$ forward-backward asymmetry. Such a measurement near 
the $Z$-pole can be sensitive to interference effects between tree-level $Z$-boson exchange and possible non-standard-model amplitudes~\cite{Z-pole}.   

The ATLAS~\cite{Z-ATLAS}, LHCb~\cite{Z-LHCb} and CMS~\cite{CMS_Hboost} collaborations have measured the $p p \rightarrow Z \rightarrow b\bar{b}$  
process at various $Z$-boson momenta and rapidities at the Large Hadron Collider. At the Tevatron the only determination of the $Z\to b\bar{b}$ decay was reported by the CDF collaboration, which determined 
$\sigma(p\bar{p} \rightarrow Z) \times \mathcal{B}(Z \rightarrow b\bar{b})$ with a relative uncertainty of $29\%$~\cite{donini}. Owing to a new, larger, and unique data set, and an analysis technique that uses data to determine the invariant-mass spectrum of the various multijet background contributions, the measurement reported in this paper improves the precision of the 
previous CDF analysis by almost a factor of two. 

Beyond its intrinsic value, the $Z\to b\bar{b}$ measurement allows a validation of the background description for the first inclusive search of the 
$p\bar{p} \rightarrow H \rightarrow b\bar{b}$ process. Such search has been recently reported by the CMS collaboration~\cite{CMS_Hboost}. Despite its large branching ratio~\cite{PDG}, the coupling of the Higgs boson to $b$ quarks has been established only recently~\cite{CMS_bb,Atlas_bb}. In addition, 
the inclusive search is sensitive to a broad class of non-standard-model contributions in the general context of resonances search. For example, new scalar particles decaying into $b\bar{b}$ 
quark pairs are foreseen in models where they act as mediators for dark matter interactions~\cite{DM1,DM2}. 


\section{II. The CDF detector}
The data used in this analysis were collected during Run II at the Tevatron with the CDF II detector~\cite{Acosta:2004yw}.
The CDF II detector was a multipurpose azimuthally-symmetric magnetic spectrometer with a large tracking volume inside a magnet solenoid. Outside the solenoid, there were sampling 
calorimeters, surrounded by the steel flux return of the solenoid and by muon detectors.
\section{III. Event selection}
Data were collected with a dedicated online event-selection (trigger) algorithm~\cite{CDF_b-trigger}, designed and optimized to select events with hadronic $b$-quark decays. 
Because of the large background from multijet production, fully hadronic triggers suffered for a high event-accept rate, which saturated the CDF data-taking bandwidth. This algorithm was implemented in the CDF II software in April 2008. 
It was the only online selection for fully hadronic $b$-jets capable of collecting data at instantaneous luminosities of up to $4 \times 10^{32}$\,cm$^{-2}$s$^{-1}$ while meeting the data-acquisition timing 
and bandwidth requirements. Data collected by this algorithm corresponds to a total of $5.4~\invfb$ of integrated luminosity.
 
The concept subtending this trigger is to use the online measurement of the impact parameter~\cite{IP} as an additional discriminating variable that allows relaxing the jet-energy thresholds, thus minimizing the sculpting 
on the dijet-mass distribution in the mass range of the $Z$ and Higgs bosons. The trigger algorithm is structured into three levels.
At the first level, at least two central ($|\eta| < 1.5$~\cite{DEF}) calorimetric energy deposits with $E_{\mathrm{T}} > 5$\,GeV and two charged particles with $p_\mathrm{T} > 2$\,GeV/$c$ must be 
reconstructed.
At level 2, jets with $E_{\mathrm{T}} > 15$\,GeV and $|\eta| < 1.0$ are reconstructed using a fixed-cone algorithm\,\cite{JetClu} with a radius, $\Delta R=\sqrt{\Delta\eta^2+\Delta\phi^2}$, 
of 0.7.  At least two tracks with impact parameter $d_0>90~\mu$m matched to one of the jets have to be identified. The distance of the secondary vertex from 
the primary vertex in the transverse plane, $R_{\mathrm{b}}$~\cite{Rb}, is required to be greater than $0.1$\,cm. At level 3, the trigger algorithm confirms level-2 requirements using offline-quality tracks and jet variables. 

After these requirements, the final output rate is 7\,Hz, well below the CDF II maximum output rate of 150\,Hz~\cite{CDF_b-trigger}. The efficiencies for $H \to b\bar{b}$ and $Z\to b\bar{b}$ signal events 
are typically of 13\% and 6\%, respectively. 

Jets are reconstructed offline using the fixed-size cone algorithm with $\Delta R=0.7$. The event selection requires 
two central ($|\eta| < 1$) jets with $E_{\mathrm{T}} > 22$\,GeV. Identification of $b$ hadrons in the event is performed with the {\sc SECVTX} algorithm~\cite{secvtx}. The algorithm looks for a vertex 
displaced from the collision point, which is likely to be produced by the decay of a $b$ hadron. The $b$-jet candidates selected by the trigger are required to contain a secondary vertex identified by the {\sc SECVTX} algorithm, and are referred to as 
the $b$-tagged trigger jet. The sample of events containing the two leading-$E_{\mathrm{T}}$ jets, of which one is the $b$-tagged trigger jet and the other is not required to be $b$-tagged (other jet) is 
referred to as the single-tagged sample. The events of the sample used to search for $b\bar b$ resonances are required to have two {\sc SECVTX} $b$-tagged jets. If more than two 
{\sc SECVTX} $b$-tagged jets are identified in an event, the two with the largest $E_{\mathrm{T}}$ are chosen. We refer 
to this sample as the double-tagged sample. The double-tagged sample contains $925\,338$ events.

\section{IV. Event simulation}

Simulated events are used to evaluate the acceptances for the $Z\rightarrow b \bar b$, and $H\rightarrow b \bar b$ signal processes, and to determine the efficiencies of the {\sc SECVTX} algorithm for the various 
jet-flavor hypotheses by exploiting $b\bar{b}$, $c\bar{c}$, and light-quark samples. 
Samples are generated using {\sc Pythia}~\cite{pythia} with the CTEQ5L~\cite{CTEQ5L} set of parton distribution functions (PDF). The different responses of the online and the offline $b$-tagging algorithms in data and 
simulated jets are reconciled by using correction scale factors. While the $b$-tagging efficiencies can be determined in simulated samples, more complex is their extraction in data, where the $b$-quark content of the 
sample is not known a priori. 

The procedure to extract the $b$-tagging efficiencies in data uses events with muons reconstructed inside the jet cone, which are likely to originate from semileptonic decays of $b$ hadrons, $B\to \mu X$. Because of the high 
mass of $b$ hadrons, the muon momentum transverse to the jet axis, $p_{\text{T,rel}}$, is typically larger than for muons from generic semileptonic hadron decays. 
A data control sample, collected using a trigger that required the presence of a muon with $p_{\text{T}} > 8$\,GeV/$c$, is exploited to that end. The fraction of $b\bar{b}$ events in this data set is 
determined by fitting the $p_{\text{T,rel}}$ distribution of muons reconstructed inside a jet cone. Then the $b$-tagging efficiencies, and therefore the factors to scale such efficiencies from simulation 
to data, are determined as functions of the jet $E_{\mathrm{T}}$. They are considered constant in the analysis, with a value of $0.86 \pm 0.01(\text{stat}) \pm 0.04(\text{syst})$ 
for the offline $b$-tagging and $0.68\pm 0.01(\text{stat}) \pm 0.03(\text{syst})$ for the trigger and offline combined $b$-tagging. The small energy dependence observed is accounted for in the systematic uncertainties. 
The systematic uncertainties of the scale factors comprise the different response of the tagging algorithm to $b$-jets with a muon with respect to fully hadronic $b$-jets and the uncertainty on the modeling of the jet 
direction. 

\section{V. Background model}
The search for $Z\rightarrow b \bar{b}$ and $H\rightarrow b \bar{b}$ candidates is conducted by looking for an enhancement over the continuum background in the invariant-mass distribution of the two 
leading-$E_{\mathrm{T}}$ $b$-tagged jets, $m_{12}$, in the double-tagged sample. This sample is predominantly composed of $b$-jet pairs from multijet production, but the fraction of events 
with one jet initiated by a charm or light quark and wrongly identified as a $b$-jet is not negligible {\it a priori}. 

The multijet background composition of the sample cannot be determined reliably from simulation, which is affected by large theoretical uncertainties. A data-driven method is used to 
evaluate and model this background contribution following Ref.~\cite{Tom_Wright}. This method allows a determination of the shapes of the invariant-mass distribution of the backgrounds from data, 
by exploiting the single-tagged sample. We refer to these shapes as background templates. The normalizations are determined with a fit to the binned mass distribution of the double-tagged sample 
and are part of the results of the analysis. The signal templates are derived from the double-tagged samples selected from simulated samples of $Z$- and Higgs-boson decays. 

In order to determine the background templates, the flavor-dependent bias introduced by the {\sc SECVTX} tagging is reproduced on the other jet of the single-tagged sample by weighting it with tagging 
probabilities for $b$-, $c$-, and light-quark jets; we refer to this jet as the simulated flavor-tagged jet. The invariant mass of the dijet formed by the $b$-tagged trigger jet and the simulated flavor-tagged jet is 
calculated under the various flavor hypotheses. 

This method allows for predictions of the shape of the background by using the single-tagged sample, which is expected to have negligible signal contamination ($0.3\%$ estimated using a simulated signal sample). 
Simulated $b\bar{b}$, $c\bar{b}$ and light-quark samples are employed only to determine the per-jet probabilities that a jet initiated by a $b$, $c$, or light quark is tagged as $b$-jet by the 
{\sc SECVTX} algorithm as functions of jet $E_{\mathrm{T}}$ and $\eta$. These probabilities are referred to as tagging matrices. 

\begin{figure}
\centering
\includegraphics[width=1\columnwidth]{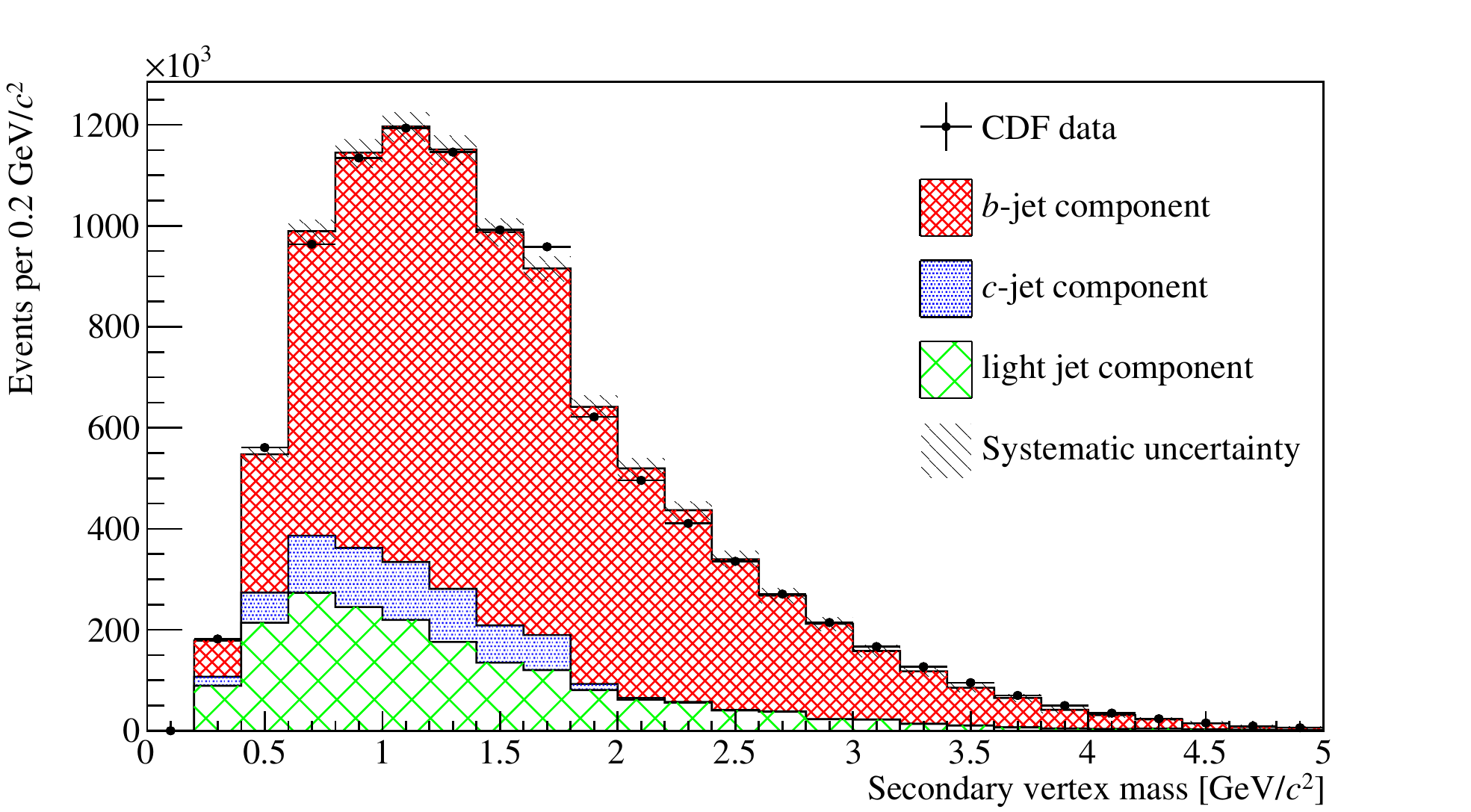}
\caption{Invariant-mass distribution of the charged particles originating from the secondary vertex, $M_{\text{SV}}$ in the text, of $b$-tagged trigger jets from the single-tagged sample with fit results 
overlaid as stacked components.}
\label{fig:TagMass}
\end{figure}

This procedure relies on the assumption that the $b$-tagged trigger jet in the single-tagged sample is initiated by a $b$ quark. The invariant mass of the secondary vertex, $M_{\text{SV}}$, is 
used to determine the heavy-flavor content of the $b$-tagged trigger jet sample. The quantity $M_{\text{SV}}$ is defined as the invariant mass of all charged particles originating from 
the secondary vertex, where all particles are assumed to be pions. This variable is sensitive to the flavor of the parton initiating the jet. Light quarks and gluons, where only track mismeasurements may 
generate a secondary vertex tag, favor low $M_{\text{SV}}$ values.
Hadrons originating from $b$ quarks have larger $M_{\text{SV}}$ values than those originating from $c$ quarks.
Figure \ref{fig:TagMass} shows the $M_\text{SV}$ distribution of $b$-tagged trigger jets from the single-tagged sample, fitted with a binned likelihood as the sum of three 
contributions: jets initiated by $b$ quarks, $c$ quarks, and light quarks. The $M_\text{SV}$ templates of the $b$-, $c$-, and light-quark jets are obtained from simulated $b\bar{b}$, $c\bar{c}$, and light-quark dijet 
samples, respectively. In the single-tagged sample, the fraction of the $b$-tagged trigger jets originating from $b$ quarks is $(75\pm 2)\%$, the fraction from $c$ quarks is $(7 \pm 1)\%$, and that from light quarks is 
$(18 \pm 2)\%$, where the uncertainties are the quadratic sum of the statistical uncertainty and the systematic uncertainty due to the finite 
size of the simulated templates. The reduced $\chi^2$ is 0.76, with 21 degrees of freedom.

By requiring $M_{\text{SV}}>1.8$\,GeV/$c^{2}$ for the $b$-tagged trigger jets in the single-tagged sample, we strongly reduce the contamination from $c$ and light quarks, thus achieving a nearly 
pure sample of $b$-tagged trigger jets originating from $b$ quarks. This sample is the starting point for the construction of the background templates. The untagged jet is 
weighted using the probabilities, functions of $E_{\mathrm{T}}$ and $\eta$, that it would be tagged by the {\sc SECVTX} algorithm if it was a jet initiated by a $b$, $c$, or light quark. The weighting probabilities 
are provided by the flavour-dependent tagging matrices. 
The background templates are the invariant-mass distributions of the dijets formed by the $b$-tagged trigger jet and the weighted untagged jet, under the three flavor hypotheses. 
The contribution of events where both tagged jets are not initiated by $b$ quarks, i.e. $c\bar{c}$ and $q\bar{q}$ dijets, is expected to be negligible and it is not considered.

The background components are classified as Bb, bB, Bc, cB, Bq and qB. Uppercase B indicates the $b$-tagged trigger jet of the single-tagged sample, while lowercase 
letters correspond to the flavor hypotheses obtained from the tagging matrices, where $q$ indicates the light quarks, applied to the untagged jet. The order of the letters follows 
the decreasing-$E_{\mathrm{T}}$ ordering of the jets. Since the dijet-mass distribution templates built with the $b$-tagging and the $c$-tagging matrices are indistinguishably similar (Bb and Bc, as well as bB and cB pairs, 
respectively), they are merged assuming a fixed $5\%$ $c$-jet contribution. A systematic uncertainty is assigned for this 
assumption. Therefore, four different templates are used to model the multijet background, Bb+Bc, bB+cB, Bq, and qB. 

\section{VI. Fitting procedure}
The dijet-mass signal templates are obtained from simulated events. The fit to the measured dijet invariant-mass distribution is performed by maximizing a binned likelihood, defined as
\begin{equation}
 \mathcal{L}=\prod_{i=1}^N  \frac{\sum_{\mathrm{s}}n_{\mathrm{s}}^i P_{\mathrm{s}}(m_{\mathrm{12}}^i) + \sum_{\mathrm{b}} n_{\mathrm{b}}^i P_{\mathrm{b}} (m_{\mathrm{12}}^i)}{\sum_{\mathrm{s}} n_{\mathrm{s}}^i + \sum_{\mathrm{b}} n_{\mathrm{b}}^i},
\end{equation}
where $\mathcal{L}$ is the product, over all bins, of the probabilities that the event in the $i$th bin of invariant mass belongs to a signal or background process. These probabilities are described by the 
background probability density functions (pdf) $P_{\mathrm{b}}(m_{\mathrm{12}})$ plus the $Z$- and the $H$-boson signal pdfs $P_{\mathrm{s}}(m_{\mathrm{12}})$. The free 
parameters are the number of signal ($n_{\mathrm{s}}^i$) and background ($n_{\mathrm{b}}^i$) events, which are constrained to be non-negative.

\begin{figure}
\centering
\includegraphics[width=\columnwidth]{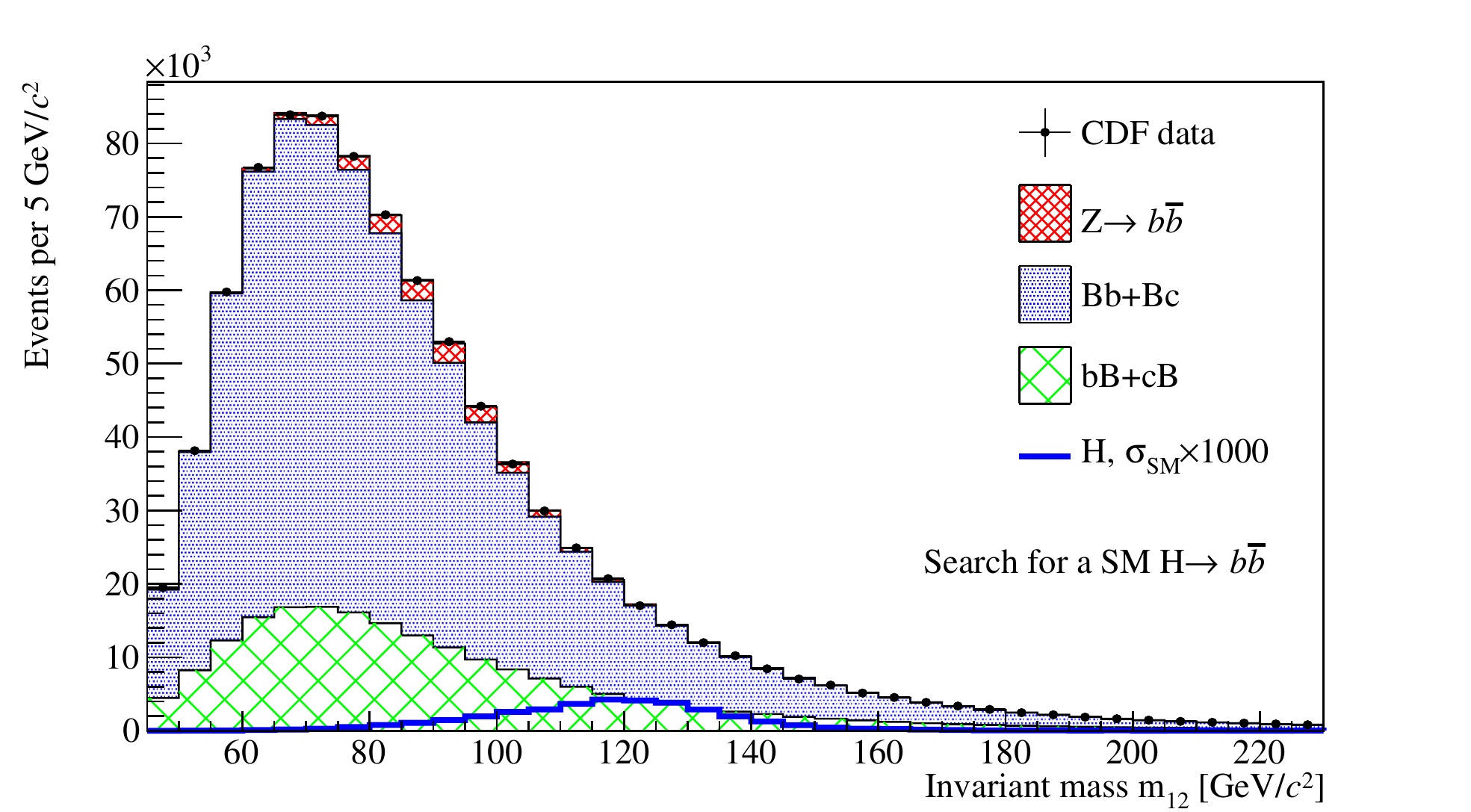}
\caption{Dijet mass distribution in double $b$-tagged events, with the results of the fit overlaid. The normalization of the Higgs-boson signal is set to $10^3$ times the expected standard-model cross section for illustrative purposes.}
\label{fig:simplefit}
\end{figure}

Figure~\ref{fig:simplefit} shows the result of the fit to the dijet-mass distribution for the double-tagged sample. The resulting yields are listed in Table~\ref{table:simplefit} where the uncertainties are statistical only.
The fit returns a sizable $Z\to b\bar{b}$ signal component, while no $H\to b\bar{b}$ signal is found. The Higgs-boson event selection efficiency is $1.5\%$, as estimated from simulation. According to the predicted 
total Higgs-boson production cross section and branching ratio into a pair of $b$ quarks, the signal acceptance, the trigger and selection efficiencies and data-to-simulation scale factors, we expect about $36$ 
Higgs-boson events in this sample.  In Figure~\ref{fig:simplefit}, the Higgs-boson component in Figure~\ref{fig:simplefit} is magnified 
$10^3$ times with respect to this expectation for illustrative purposes. The background contribution constituted by dijets where one of the two tagged jet is initiated by a light-quark component is compatible with zero, 
indicating that data are dominated by pairs of heavy-flavor jets. The reduced $\chi^2$ is 0.87, with 31 degrees of freedom. If the $Z$ and $H$ signal components 
are removed in the fit, the reduced $\chi^2$ is 7.48, with 33 degrees of freedom.
 
\begin{table}[h!]  
\centering         
 \caption{Signal and background yields determined by the fit to the double $b$-tagged data sample, with associated statistical uncertainty. If the yield is compatible with 0, the $1\sigma$ statistical uncertainty is reported.}
 \begin{ruledtabular}
 \begin{tabular}{c l}  

Component & Event yields\\ 

$H \rightarrow b\bar{b}$ & $<250~(1\sigma)$\\
$Z \rightarrow b\bar{b}$ & $(16.5 \pm  1.2) \times 10^3 $\\
Bb+Bc& $ (68.1  \pm  1.1) \times 10^4 $ \\
bB+cB & $ (19.4  \pm  1.3) \times 10^4 $ \\
Bq & $ <175~(1\sigma)$ \\
qB & $ <61~(1\sigma)$ \\

\end{tabular}  
\end{ruledtabular}
\label{table:simplefit}                                                           
\end{table}                   

From the $Z\to b\bar{b}$ yield, the product of the $Z$-boson production cross section and the branching ratio is determined using
\begin{equation}
   \sigma_Z \times \mathcal{B}(Z \rightarrow b\bar{b})=\frac{N_{\mathrm{sig}}}{\epsilon_{\mathrm{trig}} \cdot \epsilon_{\mathrm{kin}} \cdot \epsilon_{\mathrm{tag}} \cdot {\it SF}_{\mathrm{trig}} \cdot {\it SF}_{\mathrm{tag}} \cdot L},
\label{eq:CS}
\end{equation}
where $\epsilon_{\mathrm{trig}}$=6.4\%, $\epsilon_{\mathrm{kin}}$=56\%, and $\epsilon_{\mathrm{tag}}$=13\%, are the efficiencies for the trigger, offline selection and tagging requirements, respectively; 
$SF_{\mathrm{trig}}$=0.68 and $SF_{\mathrm{tag}}$=0.86 are the online and offline scale factors that match the simulated $b$-tagging performance to that on data, and $L$ is the integrated luminosity. 

The $Z\to b\bar{b}$ event sample allows for a measurement of the scale factor that matches simulated and observed calorimeter energy scales for $b$-jets. The scale factor is determined 
by fitting the $b\bar{b}$ dijet-mass distribution in data multiple times, using in each a different $Z$-signal mass template corresponding to a specific choice of the jet-energy scale. The $Z\to b\bar{b}$ 
yield is determined independently in each fit. The energy of each jet of the $Z \rightarrow b\bar{b}$ simulated sample is multiplied by a factor $k_{\text{JES}}$, which varies between 0.90 and 1.10 in steps of 0.01. This range largely 
covers the possible variation of this parameter, as established in a previous CDF analysis~\cite{donini}. The value $k_{\text{JES}}^{\text{min}}$ that minimizes the $\chi^2$ is chosen as the 
central jet-energy scale that matches simulation to data. The standard deviation on $k_{\text{JES}}^{\text{min}}$ is determined to be the $k_{\text{JES}}$ interval corresponding to a symmetric unit 
variation in $\chi^2$.

\section{VII. Systematic uncertainties}
The $Z \rightarrow b\bar{b}$ cross section and jet-energy scale measurements are affected by systematic uncertainties. 
Some of them are related to differences between data and simulation, while others are related to the signal extraction procedure. The systematic uncertainty due to the statistical fluctuations of 
the background templates is estimated by generating simplified simulated experiments that fluctuate the number of events in each bin of the background templates and measuring the resulting bias. 
The effect of the finite size of the simulated signal template is evaluated analogously, and it affects both the cross section and the jet energy scale measurement. The systematic uncertainty due to the fixed proportion of $c$-jets in the Bb and bB background templates is 
set by varying this percentage from 0 to 10\%. The data-to-simulation online and offline $b$-tagging scale factors 
are affected by systematic uncertainties that propagate to the evaluation of the signal cross section. Two separate scale factors are evaluated. One parametrizes the combined data-to-simulation 
response of the trigger and the offline $b$-tagging, and it is applied to the $b$-tagged trigger jet. The other parametrizes the offline $b$-tagging scale factor, and it is applied to the second 
$b$-tagged jet. The systematic uncertainties related to these two sources are 100\% correlated. Their energy dependence is also taken into account. The systematic 
uncertainty on the signal efficiency due to the CDF jet-energy correction is estimated by shifting the energy of simulated jets by the amount prescribed by 
the standard correction. Finally, the effect of lower or higher final-state radiation on the simulated signal is evaluated by 
generating samples with different final-state radiation tunings. The effect on the measurement of a particular choice of PDF is measured by generating samples using the CTEQ6L 
PDF set. Not all the PDF sets are considered since the impact of these variations is negligible. A summary of the systematic effects considered is
shown in Table~\ref{table:sys}. 

\begin{table}[h!]                                                                             
\centering 
 \caption{Summary of the systematic uncertainties on the $Z\to b\bar{b}$ measurements. Dashes indicate that the corresponding uncertainty does not apply.}   
 \begin{ruledtabular}
\begin{tabular}{l c c}
Source &  JES & Cross section \\ 
Integrated luminosity & $-$ & 6\% \\
Background template fluctuations & 0.4\% & 3\% \\
$c$-quark component in $b\bar{b}$ templates & 0.5\% & 2\%\\
Simulated signal sample size & 0.2\% & 3\% \\
$b$-tagging energy dependence & 0.4\% & 5\% \\
$b$-tagging scale factor & $-$ & 5\%\\
Trigger and $b$-tagging combined scale factor & $-$ &4\% \\
Jet energy correction & $-$ &1\% \\
Final-state radiation  & $-$ &$3\%$ \\  
Parton distribution functions & $-$ &$1\%$ \\ 
\hline
Total & 0.8\%& 13\% \\

\end{tabular}            
             \end{ruledtabular}                      
 \label{table:sys} 
\end{table}    

\section{VIII. Results}

The $Z$-boson production cross section times the $Z\to b\bar{b}$ branching ratio is calculated from Eq.~\eqref{eq:CS}. The measured value of 
$\sigma_Z \times \mathcal{B}(Z\rightarrow b\bar b)= 1.11\pm 0.08(\text{stat}) \pm 0.14(\text{syst})~\text{nb}$ is consistent with the next-to-leading order theoretical 
calculation~\cite{Z_teo} which predicts $\sigma_Z \times \mathcal{B}(Z\rightarrow b\bar b)= 1.13\pm 0.02~\text{nb}$. The significance of the signal 
contribution exceeds $5\sigma$, as determined using simplified simulated experiments, including statistical and systematic uncertainties. The measured $b$-jet-energy 
scale is $k_{\text{JES}}=0.993 \pm 0.022(\text{stat})\pm0.008(\text{syst})$. 

Since no significant $H\to b\bar{b}$ signal is found, a $95\%$ confidence-level (C.L.) upper limit is set to the cross section for inclusive Higgs-boson production followed by decay into a $b\bar{b}$ 
pair using the modified frequentist CL$_{\text{s}}$ method \cite{CLs,MClimit}. Simplified simulated experiments are generated based on the background description and the various assumed 
signal strengths. As a test statistic we use the distribution of the difference in $\chi^2$ between fits with a signal-plus-background model and background-only model, as resulting from fits to the 
simplified simulated experiments. The systematic uncertainties associated with the Higgs-boson search are the same as those considered for the $Z$-boson measurement. These uncertainties may affect both the 
normalization and the shape of the invariant-mass distributions. In the limit calculation they are introduced, using a Bayesian technique, as nuisance parameters. 
The posterior density is then marginalized with respect to these assuming Gaussian prior densities. 

\begin{figure}
\includegraphics[width=\columnwidth]{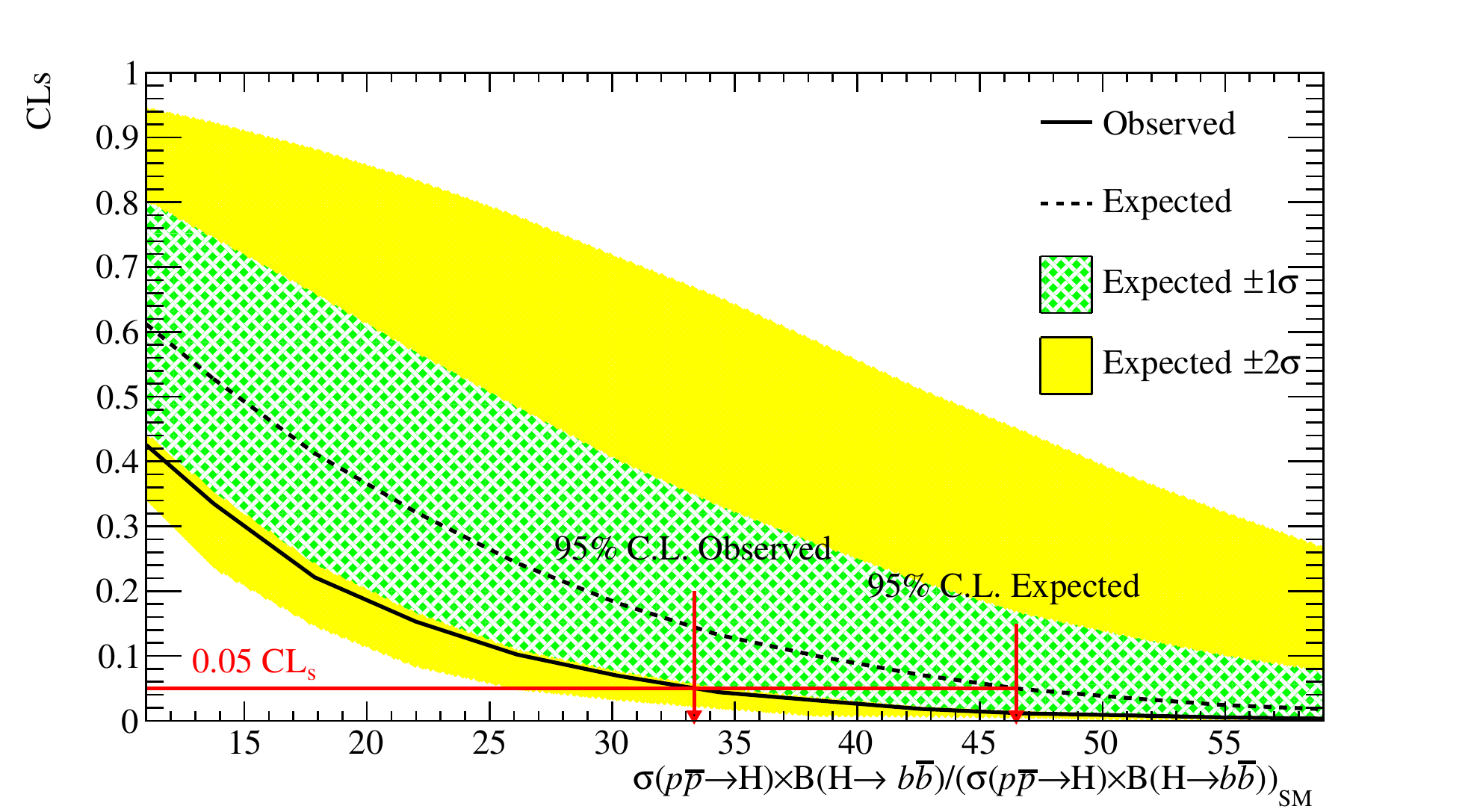}
\caption{Observed (black solid line) and expected (black dashed line) CL$_{\text{s}}$ as functions of the cross section times the branching ratio normalized to the standard-model prediction. 
Dark and light regions are the bands covering the expected $68.3\%$ C.L. and $95.5\%$ C.L. limits in the absence of signal, respectively. The 0.05 CL$_{\text{s}}$ 
level, which corresponds to the $95\%$ C.L. upper limits, is indicated by the red horizontal line.}
\label{fig:CLs}
\end{figure}

Figure~\ref{fig:CLs} shows the expected and observed CL$_{\text{s}}$ values as functions of the cross section times the branching ratio normalized to the standard-model 
$H \rightarrow b\bar{b}$ prediction. The observed (expected) upper limit at 95\% C.L. on the $p\bar p \rightarrow H\rightarrow b\bar b$ process is 33 (46) times the standard-model 
cross section, which corresponds to a cross section of $40.6$ ($56.6$)\,pb. This represents the first inclusive limit of the $p\bar p \rightarrow H\rightarrow b\bar b$ process at $\sqrt{s}=1.96$ TeV.

The measurements reported in this paper can be considered the CDF legacy for the $\sigma(p\bar{p} \rightarrow Z) \times \mathcal{B}(Z \rightarrow b\bar{b})$ determination and the inclusive 
Higgs-boson production cross section limit.

\section{Acknowledgements}
We thank the Fermilab staff and the technical staffs of the participating institutions for their vital contributions. This work was supported by the U.S. Department of Energy and National 
Science Foundation; the Italian Istituto Nazionale di Fisica Nucleare; the Ministry of Education, Culture, Sports, Science and Technology of Japan; the Natural Sciences and Engineering Research 
Council of Canada; the National Science Council of the Republic of China; the Swiss National Science Foundation; the A.P. Sloan Foundation; the Bundesministerium f\"ur Bildung und Forschung,
Germany; the Korean World Class University Program, the National Research Foundation of Korea; the Science and Technology Facilities Council and the Royal Society, UK; the Russian Foundation for 
Basic Research; the Ministerio de Ciencia e Innovaci\'{o}n, and Programa Consolider-Ingenio 2010, Spain; the Slovak R\&D Agency; the Academy of Finland; the Australian Research Council (ARC); and 
the EU community Marie Curie Fellowship contract 302103.


\begin{thebibliography}{usrt}
\bibitem{DEF} CDF uses a cylindrical coordinate system with azimuthal angle $\phi$, polar angle $\theta$ measured with respect to the positive $z$ direction along the proton beam, and distance $r$ measured from the beamline. The pseudorapidity, transverse energy, and transverse momentum are defined as $\eta=-\ln[\tan(\theta/2)]$, $E_{\mathrm{T}}=E\sin \theta$, and $p_{\mathrm{T}} = p \sin \theta$, respectively, where $E$ and $p$ are the energy and momentum of an outgoing particle.
\bibitem{IP} The tranverse impact parameter is defined as the distance in the transverse plane with respect to the beamline at the track's closest approach.
\bibitem{Rb} The $b$-hadron decay length in the transverse plane, $R_{\mathrm{b}}$, is defined in $ d_{\mathrm{0}}=R_{\mathrm{b}} \sin(\phi_\mathrm{b}-\phi)$ with $\phi$ and $\phi_\mathrm{b}$ the azimuthal angles of the jet and of the $b$-hadron, respectively.
\end{thebibliography}

\end{document}